\input harvmac.tex
\noblackbox
%


\def\unlockat{\catcode`\@=11}
\def\lockat{\catcode`\@=12}

\unlockat

\def\newsec#1{\global\advance\secno by1\message{(\the\secno. #1)}
\global\subsecno=0\global\subsubsecno=0\eqnres@t\noindent
{\bf\the\secno. #1}
\writetoca{{\secsym} {#1}}\par\nobreak\medskip\nobreak}
\global\newcount\subsecno \global\subsecno=0
\def\subsec#1{\global\advance\subsecno
by1\message{(\secsym\the\subsecno. #1)}
\ifnum\lastpenalty>9000\else\bigbreak\fi\global\subsubsecno=0
\noindent{\it\secsym\the\subsecno. #1}
\writetoca{\string\quad {\secsym\the\subsecno.} {#1}}
\par\nobreak\medskip\nobreak}
\global\newcount\subsubsecno \global\subsubsecno=0
\def\subsubsec#1{\global\advance\subsubsecno by1
\message{(\secsym\the\subsecno.\the\subsubsecno. #1)}
\ifnum\lastpenalty>9000\else\bigbreak\fi
\noindent\quad{\secsym\the\subsecno.\the\subsubsecno.}{#1}
\writetoca{\string\qquad{\secsym\the\subsecno.\the\subsubsecno.}{#1}}
\par\nobreak\medskip\nobreak}

\def\subsubseclab#1{\DefWarn#1\xdef
#1{\noexpand\hyperref{}{subsubsection}%
{\secsym\the\subsecno.\the\subsubsecno}%
{\secsym\the\subsecno.\the\subsubsecno}}%
\writedef{#1\leftbracket#1}\wrlabeL{#1=#1}}
\lockat

%
%
%

\def\IZ{\relax\ifmmode\mathchoice
{\hbox{\cmss Z\kern-.4em Z}}{\hbox{\cmss Z\kern-.4em Z}}
{\lower.9pt\hbox{\cmsss Z\kern-.4em Z}}
{\lower1.2pt\hbox{\cmsss Z\kern-.4em Z}}\else{\cmss Z\kern-.4em
Z}\fi}
\def\IB{\relax{\rm I\kern-.18em B}}
\def\IC{{\relax\hbox{$\inbar\kern-.3em{\rm C}$}}}
\def\ID{\relax{\rm I\kern-.18em D}}
\def\IE{\relax{\rm I\kern-.18em E}}
\def\IF{\relax{\rm I\kern-.18em F}}
\def\IG{\relax\hbox{$\inbar\kern-.3em{\rm G}$}}
\def\IGa{\relax\hbox{${\rm I}\kern-.18em\Gamma$}}
\def\IH{\relax{\rm I\kern-.18em H}}
\def\II{\relax{\rm I\kern-.18em I}}
\def\IK{\relax{\rm I\kern-.18em K}}
\def\IP{\relax{\rm I\kern-.18em P}}

\def\inbar{\,\vrule height1.5ex width.4pt depth0pt}

\font\cmss=cmss10 \font\cmsss=cmss10 at 7pt
\def\IR{\relax{\rm I\kern-.18em R}}

\def\sdtimes{\mathbin{\hbox{\hskip2pt\vrule
height 4.1pt depth -.3pt width .25pt\hskip-2pt$\times$}}}

\def\bZ{{\bf Z}}
\def\bR{{\bf R}}

\def\npb#1#2#3{{\sl Nucl. Phys.} {\bf B#1} (#2) #3}
\def\plb#1#2#3{{\sl Phys. Lett.} {\bf B#1} (#2) #3}
\def\prl#1#2#3{{\sl Phys. Rev. Lett. }{\bf #1} (#2) #3}
\def\prd#1#2#3{{\sl Phys. Rev. }{\bf D#1} (#2) #3}

\def\cmp#1#2#3{{\sl Comm. Math. Phys. }{\bf #1} (#2) #3}
\def\mpl#1#2#3{{\sl Mod. Phys. Lett. }{\bf #1} (#2) #3}

\lref\JuliI{B. Julia, ``Group Disintegrations,'' in {\it Superspace and
Supergravity}, Proceedings of the Nuffield Workshop, 1980, eds. S. W.
Hawking and M. Rocek, Cambridge Universtiy Press, (Cambridge 1981).}

\lref\JuliII{B. Julia, in {\it Applications of group theory in physics
and mathematical physics,} ed. P. Sally et. al. (American Mathematical
Society, Providence, 1985). }

\lref\CrJu{E. Cremmer and B. Julia, {\sl Nucl. Phys.} {\bf B159} (1979) 141.}

\lref\Town{P. Townsend, ``The Eleven-dimensional Supermembrane Revisited,''
\plb{250}{1995}{184}, hep-th/9501068.}

\lref\Vafa{C. Vafa, ``Modular Invariance and Discrete Torsion on Orbifolds,''
\npb{273}{1986}{592}.}

\lref\VafaII{C. Vafa, ``Evidence for F-theory,'' \npb{469}{1996}{403},
hep-th/9602022.}

\lref\BKMT{P. Berglund, A.~Klemm, P.~Mayr, and S.~Theisen,
``On Type IIB Vacua With Varying Coupling Constant,''
hep-th/9805189.}

\lref\FrVa{D. Freed and C. Vafa, ``Global Anomalies on Orbifolds,''
\cmp{110}{1987}{349}.}

\lref\NSV{K. Narain, M. Sarmadi, and C. Vafa, ``Asymmetric Orbifolds,''
\npb{288}{1987}{551}.}

\lref\DKV{L. Dixon, V. Kaplunovsky, and C. Vafa, ``On Four-dimensional gauge
Theories from Type-II Superstrings,'' \npb{294}{1987}{43}.}

\lref\ADF{L. Adrianopoli, R. D' Auria, and  and S. Ferrara, ``U-duality
and Central Charges in Various Dimensions Revisited'',
{\sl Int. J. Mod. Phys.} {\bf A13} (1998) 431, hep-th/9612105.}

\lref\WittI{E. Witten, ``String Theory Dynamics in Various Dimensions,''
Nucl. Phys. {\bf B443} (1995) 85, hep-th/9503124.}

\lref\WittII{E.~Witten, ``Is Supersymmetry Really Broken,''
{\sl Int. J. Mod. Phys.}{\bf A10} (1995) 1247, hep-th/9409111.}

\lref\WittIII{E.~Witten, ``Strong Coupling and the Cosmological
constant,'' \mpl{A10}{2153}{1995}, hep-th/9506101.}

\lref\BFSS{T. Banks, W. Fischler, S. Shenker and L. Susskind.
``M Theory as a Matrix Model: A Conjecture,'' \prd{55}{1997}{},
hep-th/9610043.}

\lref\Bank{T. Banks,``Matrix Theory,'' {\sl Nucl. Phys. Proc. Suppl.}
{\bf 67} (1998) 180, hep-th/9710231.}

\lref\BiSu{D. Bigatti and L. Susskind, ``Review of Matrix Theory,''
hep-th/9712072.}

\lref\Nara{K. S. Narain, ``New Heterotic String Theories in Uncompactified
Dimension $< 10$,'' { \sl Phys. Lett.} {\bf 169B} (1986) 41.}

\lref\Bouw{P. Bouwknegt, ``Lie Algebra Automorphisms, the Weyl Group,
and Tables of Shift Vectors,'' {\sl J. Math. Phys.} {\bf 30(3)} (1989) 571.}

\lref\Cart{R. W. Carter, ``Conjugacy Classes in the Weyl Group,''
{\sl Comp. Math.} {\bf 25} (1972) 1.}

\lref\Lerc{W. Lerche, A. N. Schellekens and N. P. Warner,
``Lattices and Strings,'' {\sl Phys. Rept.} {\bf 177} (1989) 1.}

\lref\VaWi{C. Vafa and E. Witten, Dual String Pairs with $N=1$ and
$N=2$ Supersymmetry in Four Dimensions, \sl{Nucl. Phys. Proc. Suppl.}
{bf46} (1996) 225, hep-th/9507050.}

\lref\SeVa{A. Sen and C. Vafa, ``Dual Pairs of Type-II String
Compactification, \npb{455}{1995}{165}, hep-th/9508064.}

\lref\SGM{A. Strominger, B. Greene and D. Morrison, ``Black Hole
Condensation and the Unification of String Vacua,'' {\sl Nucl. Phys.}
{\bf B451} (1995) 109, hep-th/9504145.}

\lref\SaSe{{\it Supergravities in Diverse Dimensions,} Eds. A. Salam
and E. Sezgin, North Holland and World Scientific, 1989.}

\lref\AlWi{L. Alvarez-Gaum\'e and E. Witten, ``Gravitational Anomalies,''
\npb{234}{1984}{269}.}

\lref\WiGlob{E. Witten, ``Global Gravitational Anomalies,'' Commun. Math.
Phys. {\sl Commun. Math. Phys.} {\bf 100} (1985) 197.}

\lref\GrSc{M. Green and J. Schwarz, ``Anomaly Cancellations
in Supersymmetric $ D=10$ Gauge Theory and Superstring Theory,''
\plb{149}{1984}{117.}}

\lref\esref{Ref on $N=7$ becoming $N=8$}

\lref\DaHa{A. Dabholkar and J. Harvey,  {\it in progress},}

\lref\HMV{J. A. Harvey. G. Moore and C. Vafa, ``Quasicrystalline
Compactification,''  \npb{304}{1988}{269}.}

\lref\CHL{S. Chaudhuri, G. Hockney, and J. Lykken, ``Maximally
Supersymmetric String Theories in $ D<10$,''
\prl{75}{1995}{2264}, hep-th/9505054.}

\lref\ds{M. Dine and N. Seiberg, ``Couplings and Scales in Superstring
Models,'' Phys. Rev. Lett. {\bf 55} (1985) 366.}

\lref\kks{S. Kachru, J. Kumar and E. Silverstein, ``Vacumm Energy Cancellation
in a Non-supersymmetric String,'' hep-th/9807076.}

\lref\ginsparg{P. Ginsparg, ``Curiosities at $c=1$,'' Nucl. Phys.
{\bf B295} (1988) 153.}

\lref\dinesil{M. Dine and E. Silverstein, ``New M-theory Backgrounds
with Frozen Moduli,'' hep-th/9712166.}

\lref\lanced{This example is due to L. Dixon, unpublished.}

\lref\muwit{M. Mueller and E. Witten, ``Twisting Toroidally Compactified
Heterotic Strings with Enlarged Symmetry Groups,'' {\sl Phys. Lett.}
{\bf 182B} (1986) 28.}

\lref\cremmer{E. Cremmer in ``{\it Cambridge 1980, Proceedings, Superspace
and Supergravity},'' eds. S. W. Hawking and M. Rocek, Cambridge
University Press, 1981.}

\lref\chamnic{A. H. Chamseddine and H. Nicolai, {\sl Phys. Lett.}
{\bf 96B} (1980) 89.}

\lref\mizohta{S. Mizoguchi and N. Ohta, ``More on the Similarity between
$D=5$ Simple Supergravity and M Theory,'' hep-th/9807111.}

\lref\banksrev{For a review see T. Banks, ``SUSY Breaking, Cosmology,
Vacuum Selection and the Cosmological Constant in String Theory,''
hep-th/9601151.}

\lref\dvv{R. Dijkgraaf, E. Verlinde and H. Verlinde, ``$c=1$ Conformal
Field Theory on Riemann Surfaces,'' Comm. Math. Phys. {\bf 115} (1988)
649.}

\lref\atkinl{G. Moore, ``Atkin-Lehner Symmetry,'' {\sl Nucl. Phys.}
{\bf B293}
(1987) 139., Erratum ibid. B299 (1988) 847.}

\lref\chlowe{S. Chaudhuri and D. A. Lowe, ``Monstrous String-String
Duality,'' {\sl Nucl. Phys. } {\bf B469} (1996) 21, hep-th/9512226.}
%
%

\Title{\vbox{\baselineskip12pt
\hbox{hep-th/9809122}
\hbox{EFI-98-34}
\hbox{TIFR/TH/98-30}
}}
{\vbox{\centerline{String Islands}}}
\centerline{Atish Dabholkar}
\bigskip
\centerline{\sl Department of Theoretical Physics}
\centerline{\sl Tata Institute of Fundamental Research}
\centerline{\sl Homi Bhabha Road, Mumbai, India 400005}
\bigskip
\centerline{Jeffrey A. Harvey}
\bigskip
\centerline{\sl Enrico Fermi Institute and Department of Physics}
\centerline{\sl University of Chicago, 5640 Ellis Avenue}
\centerline{\sl  Chicago, IL 60637 U.S.A.}
\bigskip

\bigskip
\centerline{\bf Abstract}

We discuss string theories with small numbers of non-compact moduli
and describe constructions of string theories whose low-energy
limit is described by various pure supergravity theories. 

\Date{September 1998}
%

\newsec{Introduction}
One of the
main obstacles preventing more direct phenomenological applications of string
theory is the problem of vacuum degeneracy. One aspect of this problem
is the presence of massless scalar fields or
moduli in compactifications of string theory.
These moduli  govern the shape and size of the compactification space as well
as the value of the coupling constant in string theory and correspond
to massless fields in spacetime. There are
stringent constraints on the presence of such massless scalars in the real
world,
so it is usually assumed that masses are generated for moduli fields
by whatever mechanism
breaks supersymmetry in string theory.
A  related problem is that the moduli fields, particularly
the dilaton which governs the value of the coupling constant, tend to
run off to infinity in known mechanisms for supersymmetry breaking,
leaving one with no vacuum at all except at zero coupling \ds.
Even in a cosmological situation the presence of moduli is
problematic \banksrev. Thus it is interesting to consider string
theories with few or no moduli.

Another reason why theories with few moduli are of
interest has to do with speculative proposals for a solution to the
cosmological constant problem \refs{\atkinl,\WittIII, \kks}. In
\atkinl\ it was suggested that the one-loop contribution to the
cosmological constant might vanish in certain special theories
where the left and right-moving contributions are
chiral with respect to an Atkin-Lehner symmetry. This naturally leads
to asymmetric orbifold constructions and hence theories with a
reduced number of moduli. In \WittIII\
it was proposed that
the cosmological constant could vanish if our four dimensional world
arises as a strong coupling limit of a three-dimensional world. In three
dimensions supersymmetry can enforce a vanishing cosmological constant
without imposing degeneracy between fermion and boson masses \WittII.
It seems quite likely that this mechanism, if it works at all, could only
work in a theory which cannot be continuously connected to higher
dimensional theories by varying moduli fields. This is because one could
first go to a higher dimensional theory and then take the strong
coupling limit which is known in many cases not to lead to
supersymmetry breaking.
Therefore one is
again interested in theories which are free of moduli other than the
dilaton or at least free of non-compact
geometrical moduli which take one to
higher dimensions. For apparently different reasons small numbers of moduli
also entered into the proposal of \kks\ based on the AdS/CFT
correspondence.  In this case one wants the theory to contain
Reissner-Nordstrom black holes with $AdS_2$  near horizon geometry.
The $AdS_2$ symmetry is spoiled by the presence of moduli
which couple to the gauge fields under which the black hole
is charged so again one is interested in theories with few moduli.

One final motivation for studying such theories is to improve our
understanding of the moduli space of string compactifications with
given spacetime supersymmetry. Of particular interest is the question
of whether this moduli space is connected. In \SGM\ it was pointed
out that many Calabi-Yau vacua which were previously thought to
be disconnected are in fact related via conifold transitions.
More generally, if we take the size of the manifold of compactification
to be very large, then locally the physics looks ten dimensional.
We will give examples here of string vacua which appear to be
isolated and are not continuously connected via vacuum
configurations to geometrical
compactifications of string theory. These theories have no
geometrical moduli and moreover
are self-dual so that even strong coupling does
not relate them to higher-dimensional theories. We refer to these vacua
as ``String Islands.'' They are higher-dimensional
supersymmetric versions of the
``Ginsparg Archipelago'' of $c=1$ CFT \ginsparg.

The basic idea behind our construction has been known for
a long time: one can remove moduli by twisting a theory by
a symmetry which exists only at special values of the moduli \muwit.
There is a great deal of literature on this subject; the novelty
here as far as we know is that we construct theories with
no moduli other than the dilaton and use string duality to
argue that the strong coupling limit does not take one back
to a higher-dimensional theory. Thus one can argue that these
theories are truly isolated in that no variation of the moduli
connects them to higher dimensional theories. Similar constructions
appear in \chlowe\ in the context of two-dimensional asymmetric
orbifolds where it was argued that varying certain radial moduli
would lead to pure supergravity theories in higher dimensions. 
In such constructions one must ensure that no new moduli appear as
some radii are taken to infinity. 
Some of the theories we
construct do have a number of moduli, but many of these
are compact and so do not take us back to a higher-dimensional
theory.

It is also interesting to ask about theories which contain no
moduli at all, not even the dilaton. Such theories exist in
compactifications to two dimensions \HMV\ and of course the most
famous example of such a theory is M theory which arises
as the strong coupling limit of IIA string theory. M theory
is described at low-energies by eleven-dimensional supergravity.
There is by now
good evidence that M theory exists beyond the low-energy
approximation  and there  is a  proposal for a more
complete formulation
of M theory \BFSS\ which has passed a number of non-trivial tests
\refs{\Bank, \BiSu}. We will comment at the end of this paper
on the possibility of other theories without any moduli at all.
For most of the paper we focus on perturbative string constructions
in dimensions four or greater which therefore always contain a dilaton.
The possibility of constructing moduli free theories by modding
out simultaneously by $S$ and $T$ dualities and some non-perturbative
aspects of theories with few moduli have been discussed
in \dinesil\ .

In searching for theories with few moduli it is not completely
clear what criteria to impose. The simplest and  most stringent
compatible with supersymmetry and the one we will impose in this
paper is to try to obtain pure supergravity
theories without matter fields as the low-energy limit of
string constructions. Many pure supergravity theories are inconsistent
because of anomalies. Of the non-anomalous pure supergravity theories
we have been able to obtain all via string constructions except
for one in $D=8$ and one in $D=7$.

In the next section we summarize
the consistent pure supergravity theories and a few of their
properties. The following section contains a set of explicit constructions
of most of these theories. We end with some general comments and
speculations.

\newsec{Consistent Pure Supergravities}

We will refer to a supergravity theory without matter multiplets as
pure supergravity. Some pure supergravities of course allow the
addition of matter multiplets while others (such as $D=11$ supergravity)
do not, but we will not make any distinction between these two
cases in what follows.
Pure supergravity theories in $D$ dimensions are uniquely specified
by the number of supersymmetries $N$. In six and ten dimensions
we must specify the number of supersymmetries of each chirality
$[N_L,N_R]$. However there does not seem to be
a uniform convention for counting supersymmetries because
of various reality conditions. We will therefore denote
theories by $D$, $N$, and $N_s$ with $N_s$ the number of
component real supersymmetries, so for example the low-energy
limit of M-theory is denoted by $(11,1,32)$. A useful reference
on the possible supergravity theories is \SaSe.

All supergravities in $D<11$ dimensions with $N_s=32$ arise
from toroidal compactification of M theory and therefore have
moduli which take one back to eleven dimensions \WittI.

All pure supergravity theories with $N_s < 16$ and
$D \ge 4$ are either
inconsistent due to anomalies ( $(6,[1,0],8)$ ) or have no
scalars at all in the supergravity multiplet ( $(5,1,8)$,
$(4,3,12)$, $(4,2,8)$, $(4,1,4)$ ). The latter theories are
interesting since they could in principle lead to lower-dimensional
theories with no moduli but because they have no dilaton they
cannot be obtained from perturbative string constructions.

Of the remaining pure supergravities in $D \ge 4$,
$(10,[1,0],16)$, $(6,[2,1],24)$ and $(6,[2,0],16)$ have
perturbative gravitational anomalies and so are inconsistent.
The $(9,1,16)$ theory is also inconsistent due to a global
gravitational anomaly. To see this note first that a single
Majorana spinor in $D=8k+1$ dimensions has a global gravitational
anomaly \refs{\AlWi,\WiGlob}.
On the other hand, we know that reduction of
the $D=10$, $(1,0)$ heterotic theory with $496$ vector
multiplets on $S^1$ gives a $D=9$, $N=1$ theory with $497$
gauge supermultiplets, each containing a single Majorana
fermion. Since this theory is consistent, it must be the
case that the pure supergravity theory has a global gravitational
anomaly which cancels the anomaly of the odd number of Majorana
fermions in the gauge supermultiplets.

We are then left with the following list of consistent pure
supergravities in $D \ge 4$ which have at least one scalar
in the supergravity multiplet and cannot be obtained from
toroidal compactification of M theory:
\eqn\sugralist{\eqalign{ (D,N,N_s)=&
(8,1,16), (7,1,16), (6,[1,1],16), (5,3,24), \cr
& (5,2,16), (4,6,24), (4,5,20), (4,4,16).\cr
}}

One family of
theories consists of the five theories
with $16$ supercharges in $D=4,5,6,7,8$. The $D=5,6,7,8$ theories all
have only a single real scalar field in the supergravity multiplet and
thus in perturbative string theory can only arise in theories where
all geometrical moduli are frozen and the dilaton is the
single scalar in the supergravity multiplet.  The $D=4$ theory has two scalars
which comprise the dilaton/axion and thus we expect the same to be
true for this theory. A reduction on $S^1$ of these theories in D dimensions
yields the $D-1$ theory with $N_s=16$ with a single vector multiplet
in addition to the supergravity multiplet.

A second family consists of the two theories with $24$ supercharges,
the $D=5,N=3$ theory and the $D=4,N=6$ theory. The $D=5,N=3$ theory contains
$14$ scalars, two of which are non-compact while the $D=4,N=6$
theory contains $30$ scalars, three of which are non-compact.
These theories are
apparently related by  reduction on an $S^1$ since the field content of the
$D=4$ theory is the same as the reduction on $S^1$ of the $D=5,N=3$
theory. This suggests that it should be possible to obtain the
$D=5,N=3$ theory as a limit of the $D=4,N=6$
theory, either by varying a geometrical modulus or by going
to strong coupling.

The $D=4,N=5$ theory is the odd man out, being  the only theory with $20$
supercharges. The supergravity multiplet contains ten scalars, but
only one of these is non-compact so again all moduli which would
correspond to going to large radius of the internal space have
been frozen.

The structure of the moduli space for these theories will be discussed
in more detail in the following section in the context of explicit
string constructions.

\newsec{Explicit Constructions}

We now turn to explicit
constructions of string theories with pure supergravity theories
as their low-energy limits.
We will construct these
theories using asymmetric orbifold constructions, although there are
undoubtedly other constructions of some of these theories
involving free fermions,  tensor
products of minimal models, or orientifolds.

First we recall a few basic facts about asymmetric orbifold
constructions \NSV.
In this paper we deal primarily
with orbifolds of Type-II string theory.
We thus start with a toroidal compactification of Type-II string
on a d-dimensional torus. At  special points in the Narain moduli space of
such compactifications we can obtain theories with purely left or right
moving symmetries. These occur at points in the Narain moduli space
where some of the T-duality symmetries have fixed points.
One very useful construction of such special points
proceeds as follows \NSV.  We choose a simply laced Lie algebra
${\cal G}$ of rank $d$ and define the lattice
$\Gamma^{d,d}({\cal G})$ \Nara\
as
\eqn\gamdef{\Gamma^{d,d}({\cal G}) = \{(p_L,p_R) \}, \qquad p_L,p_R
\in \Lambda_W({\cal G}), \qquad p_L - p_R \in \Lambda_R({\cal G}) }
with $\Lambda_R$ and $\Lambda_W$ the root and weight lattices of
${\cal G}$ respectively.  The resulting theory has purely left (and right)
moving symmetries given by elements of the Weyl group
${\cal W}({\cal G})$,
and in order to act correctly on fermion fields these must also be
elements of  $Spin(d)$. The general transformation is thus of the form
\eqn\asform{|p_L,p_R \rangle \rightarrow e^{2 \pi i (p_L \cdot v_L
- p_R \cdot v_R)} |g_L p_L, g_R p_R \rangle }
where $v=(v_L,v_R)$ is a shift vector and $g_L$ and $g_R$
lie in the intersection of ${\cal W}({\cal G}) $ with $Spin(d)$.
The right and left-moving fermions must also be twisted by
$g_{R,L}$ in order to preserve the world-sheet supersymmetry.

Classifying such orbifolds amounts to classifying conjugacy classes
of the Weyl groups of rank $d$ simply laced Lie algebras and then
for elements of each conjugacy class determining the allowed shift
vectors which are consistent with modular invariance.  In what
follows we will use the classification of and notation for Weyl group
conjugacy classes developed by Carter \Cart. Useful tables of
these conjugacy classes can be found in \refs{\Bouw, \Lerc}.
We will often
try to choose shift vectors which give positive vacuum energy in the twisted
sectors in order to ensure that there are no massless states coming
from the twisted sectors. In doing this it is important to remember
that in twisted sectors the momenta live in the lattice $I^*$ which
is dual to the lattice left invariant by the twist. Thus we will
want to choose shift vectors which are not in $I^*$.
A necessary condition for modular invariance of abelian orbifolds
is level-matching \Vafa. For $\bZ_n$ orbifolds,  level-matching
is ensured if in every sector there are states for which
$ n (E_R -E_L) = 0\, {\rm mod}\, 1$ where $E_{L,R}$ are the
left and right-moving energies.  Equivalently, there
must be physical states in every
twisted sector for every ground state of definite momentum and
winding.
This condition  is known to be sufficient for modular invariance
at one loop \Vafa. In fact it suffices
to check level matching for the ground state rather than for
all momentum and winding states and to check a single
mod  2 condition for elements of even order $2n$:
\eqn\modtcon{ p g^n p = 0 \qquad {\rm mod} ~ 2, }
for all $p \in \Gamma^{d,d}$.
In what follows we will mostly use odd order twists for
which it suffices
to check level matching for the ground state.

Conditions for modular invariance at higher loops have been
analyzed in \FrVa\ where it is shown that one-loop level matching
is not sufficient to guarantee level matching for theories with
a non-Abelian point group. Our theories all have  Abelian point
groups and so we do not expect any problems with higher loop modular
invariance, but it is not clear to us that the analysis of \FrVa\
guarantees higher loop modular invariance for abelian asymmetric
orbifolds which do not have a simple fermionic description.

As discussed above we will focus on
consistent minimal supergravities with $16 \le N_s < 32$.
Theories with $N_s=16$ occur in dimensions
$D=4,5,6,7,8$. These theories
contain $8 (D-2)$ physical boson and fermion degrees
of freedom and have half of the
maximal supersymmetry.
In Type-II string, $16$ supercharges come from the left-movers
and $16$ come from the right-movers.
This suggests that one can construct
an asymmetric orbifold with a left-moving twist which lies in
$O(10-D)$ but not in a subgroup and which thus breaks half of the
spacetime supersymmetry coming from the left-movers and removes
all Narain moduli from the untwisted sector. If we accompany
the twist by a  right-moving
shift which prevents the occurrence of massless states in the
twisted sector we clearly get the minimal supergravity spectrum for
theories with $N_s=16$.

If the left-moving twist lies in $SU(3)$ or in $SU(2)$, then
four or eight left-moving supersymmetries are preserved respectively,
giving us $20$ or $24$ supersymmetries in all. In some
examples that we discuss below, additional supersymmetries come
from the twisted sector if the the twists are not
accompanied by any shifts.

As we move down in dimension there are both more moduli that
must be projected out by the orbifold and larger orbifold groups and
more choices of shift vector
that can be used in the construction. It is not obvious a priori,
but it will turn out that going down in dimension in fact makes
it easier to remove the moduli and that as we move up in dimension
it becomes more difficult until we reach $D=7$ and $D=8$ where it
seems unlikely that asymmetric orbifold constructions of the
pure supergravity theories exist. This will become clearer as
we proceed.

We now consider such orbifolds on a case by case basis.
In what follows we write the roots of $A_n$ in an $n+1$ dimensional
orthonormal
basis $\{e_i, i = 1 \ldots n+1 \} $ as $\{ (e_i - e_j) \}$.
For $D_n$ we write the roots in terms of  $\{e_i, i = 1 \ldots n \}$
as $\{ \pm e_i \pm e_j \}$.

\subsec{$ D=4$}

\subsubsec{$ N=4$}
In order to implement the orbifold discussed above we need
a rank six simply laced Lie algebra and an element of the
Weyl group of this Lie algebra which lies in $SO(6)$ but not in a subgroup.
An inspection
of the list of conjugacy classes of Weyl groups of simply
laced Lie algebras \refs{\Bouw, \Lerc}
leads to many possibilities. Let
$\omega$ denote a primitive nth root of unity when the Weyl
group element has order $n$.
We find for
example that  $E_6$ has a conjugacy class $E_6(a_1)$
of elements of order $9$ with eigenvalues $\omega, \omega^2, \omega^4$;
$D_6$ has a
conjugacy class $D_6(a_2)$ of elements of order $6$ with eigenvalues $\omega,
\omega^3, \omega^5$;  $A_4 \times A_2$ has
a conjugacy class $A_4 \times A_2$ of elements of order $15$
with eigenvalues $\omega^3,\omega^5,\omega^6$ along with many
other possibilities.

{\bf a.} We can construct a $D=4,N=4$ model based on the conjugacy
class $E_6(a_1)$ of the $E_6$ Weyl group \lanced. We start with the
lattice
$\Gamma^{6, 6} (E_6)$. We then twist by a $Z_9$ element with
$g_R=1$ and $g_L = (\omega, \omega^2, \omega^4)$. We accompany this
by a shift with $v_L=0$ and
\eqn\eshift{v_R = {1\over 9} (1,1,-2;1,1,-2;-1,-1,+2)}
Here we have written the shift vector in terms of the
embedding of $A_2^3 \subset E_6$.
The action of the twist $g_L$ can be represented by
a $3\times 3$ matrix action
on the $A_2\times A_2\times A_2$ planes as
\eqn\nineaction{\pmatrix{
0 & 1 & 0 \cr
0 & 0 & 1 \cr
\alpha & 0 & 0 \cr
},}
where $ \alpha$ is a rotation in a single  $A_2$ plane by $2\pi/3$.

In this model $I = (0,\Lambda_R(E_6))$,
$I^* = (0,\Lambda_W(E_6))$ and we can check that
 $v_R$ is not in $I^*$,
and moreover  $3  v_R$  is also not in $I^*$  because
$3  v_R$ is in the $({\bf 3}, {\bf 3 }, {\bf \bar{3}})$ conjugacy class of
$A_2^3$,
whereas $\Lambda_W(E_6)$  has only the
$({\bf 3}, {\bf 3}, {\bf 3})$ and $({\bf \bar{3}}, {\bf \bar{3}}, {\bf
\bar{3}})$
conjugacy classes.

{\bf b.}
It is also possible to construct
a heterotic string theory with pure $D=4, N=4$ supergravity as its
low-energy limit using a free fermion construction as discussed
in \CHL.

In both of these examples one obtains  the $N=4$ pure supergravity
multiplet that contains the graviton, four gravitini, six graviphotons,
and two scalars. The only noncompact scalar is the  dilaton $\phi$
which is accompanied by the  axion $b$ that comes from the dualized
antisymmetric tensor $B_{\mu \nu}$. Together, the scalars parameterize
the coset
$SL(2, \bR)/SO(2)$. The duality group is expected to be
a subgroup of $SL(2, \bZ)$. Precise determination of this duality
group in each example is an important and interesting problem.

\subsubsec{$ N=5$}

{\bf a.} We start with a  toroidal compactification of
type II string theory
described by the lattice $\Gamma^{6,6}(A_6)$ and twist by the $\bZ_7$
symmetry generated by
\eqn\sevtwist{\eqalign{g_L & =(\omega, \omega^2, \omega^4), \qquad
                                    v_L=0 \cr
                       g_R & = 1, \qquad v_R = {1 \over 7}
                                                  (1,2,-3,0,0,0,0). \cr }}
In terms of
orthonormal basis vectors
$e_i, i=1,\ldots,7$ in $\bR^7$ the weight lattice of $A_6$ is in
the hyperplane in  $\bR^7$ that is orthogonal to $\sum_i e_i$.
The Weyl group of $A_6$ is the permutation group  $S_7$ which permutes
the seven basis vectors. A $\bZ_7$ subgroup of the Weyl group
is generated by $e_i \rightarrow e_{i+1}$ for all $i$. There are seven
eigenvalues given by the different seventh roots of unity.
The eigenvector corresponding to eigenvalue $1$  is
obviously $\sum_i e_i$ which is orthogonal to  the weight lattice.
Therefore, we can choose a complex basis in which the remaining six eigenvalues
are $\omega, \omega^2, \omega^4$ and their complex conjugates.
It is also clear that  $7 v_R$ is in the root lattice of $A_6$
and that $v_R$ is not in $I^* = \Lambda_W(A_6)$.

{\bf b.}
Another possibility is to consider the lattice $\Gamma^{6,6}(A_2^3)$
and twist by an asymmetric $\bZ_3$ symmetry
generated by
\eqn\threetwist{\eqalign{g_L & =(\omega, \omega, \omega), \qquad
                                    v_L=0 \cr
                       g_R & = 1, \qquad v_R = {1 \over 3}
                                                   (1,-1, 0; 1,-1, 0; 2, -2,
0).\cr }}

In both these models  there is no twist on the right, so all
four supersymmetries from the right are preserved. On the left, the twist is in
$SU(3)$, so one supersymmetry from the left is preserved
and together one obtains
$N=5$ supersymmetry.

In the NS-NS sector we obtain the metric $g_{\mu \nu}$, an antisymmetric
tensor $B_{\mu \nu}$ which can be dualized to a scalar $a$,
the dilaton $\phi$, and six vector fields. In the R-R sector we
find eight additional scalars and four more vector fields. Thus the moduli
consist of one non-compact scalar, the dilaton, and nine compact scalars.

The supergravity action is known to have $SU(1, 5)$ symmetry.
The ten vector fields transform in the $\bf 10$ (self-dual antisymmetric
rank-three tensor) of $SU(1, 5)$.
The classical moduli space parametrized by the ten scalars
is locally the coset
$SU(1, 5)/U(5)$.

Since there  are no states from the twisted
sectors, one can determine the  symmetry group purely from
group theory.
Our starting point is Type-II theory compactified on a 6-torus.
The moduli space is $E_{7}({\bZ})\backslash E_{7}(\bR)/SU(8)$
where the $E_7$ is in the maximally split form.
The  duality group $E_7({\bf Z})$ maps a generic point $x$ of the coset
 $E_{7}(\bR)/SU(8)$ to some other point $x'$ of the coset.
If a subgroup
$G_x$ of $E_7({\bf Z})$ leaves $x$ invariant  then it would be
a  symmetry
of the theory at $x$. One can then orbifold the theory at $x$ with the orbifold
group  $G_x$
to obtain a new theory.
$G_x$ is obviously a discrete
subgroup of  the isotropy group $SU(8)$. In our case because
the left-moving twist preserves four supersymmetries, it is a subgroup also
of $SU(3)$.
In the absence of twisted
states, we expect that the symmetry group of the
orbifold theory will be
the subgroup of
$E_{7}({\bf R})$ that commutes with $SU(3)$.
It is somewhat subtle to see that one obtains the correct real form
 $SU(1, 5)$ and not, say,  $SU(2, 4)$.
One useful observation is that,
in $SU(8)$, which is
the maximal compact subgroup
of $E_{7}$,
the centralizer of $SU(3)$
is $U(5)$. The symmetry group therefore must contain
$U(5)$ as a compact subgroup.
If we decomose the adjoint representation
of $E_7$ in terms of $SU(8) \supset SU(3)\times U(5)$ representations,
and keep only those states that are invariant under the $SU(3)$, then
the remaining $U(5)$ representations properly combine into $SU(1, 5)$
adjoint representation. Using these facts
one can easily determine, in agreement with
the supergravity considerations, that the symmetry
group is indeed $SU(1, 5)$.

One can ask what the quantum moduli space is. Supersymmetry
prevents any quantum corrections, so the question is really
only about global identifications or the U-duality group.
At the level of group theory  it seems natural to conjecture
that the U-duality group is $SU(1, 5, \bZ)$.
Physically this may not be true because, in general,
duality does not commute with orbifolding.
One supporting piece of evidence is that the quantized
electric and magnetic charges certainly
exist in the theory.  They transform as $ \bf 20$ of $ SU(1, 5)$
and satisfy the Dirac quantization condition. One way to define
the integral form $ SU(1, 5, \bZ)$ is to note that the $ \bf 20$
representation of $ SU(1, 5)$ is symplectic. The group
$ Sp(20, \bZ)$ has a natural action on the  lattice of
electric and magnetic charges.
One might therefore
define  $ SU(1, 5, \bZ)$ as the intersection of
$ Sp(20, \bZ)$ and $ SU(1, 5)$.
This conjecture can be tested by analyzing the  spectrum of
dyonic bound states in the theory.

\subsubsec{N=6}
We know two ways to obtain this theory.

{\bf a.} The first is closely related to a model discussed in
 \SeVa\ and is obtained by a $\bZ_2$ asymmetric twist on
$\Gamma^{4,4}(D_4)$ \foot{The model discussed in \SeVa\ had the same orbifold
action but started with the lattice $\Gamma^{4,4}(A_1^4)$. As pointed out
to us by E. Silverstein, this model is not modular invariant because
it does not satisfy the mod 2 condition \modtcon.}.
The $\bZ_2$ acts as $-1$ on the four left-moving coordinates of
$\Gamma^{4,4}$, as a shift by half a lattice vector on the right, and
this action is accompanied
by an asymmetric shift along one component of a $\Gamma^{2,2}$. This
theory has a limit where the radius of the untwisted and unshifted
direction goes to infinity and in this limit it gives the $D=5$, $N=3$
theory. The local moduli space of the $D=5, N=3$ theory is
\eqn\dfntm{{SU^*(6) \over USp(6)}}
The theory thus has two non-compact moduli which in this construction
correspond to the dilaton and
a modulus which is the radius of the
shifted $S^1$. Since states odd under the twist and with odd momenta
on this $S^1$ are physical, by going to infinite radius we recover
the $D=6$, $N=(2,2)$ theory and we can perturb all the way
back to $D=10$.

{\bf b.}
The second construction is the theory with an asymmetric $\bZ_3$ twist.
We take a lattice $\Gamma^{6, 6}(A_2^3)$
and twist by
\eqn\sometwist{\eqalign{g_L & =(\omega, \omega, \omega), \qquad
                                    v_L=0 \cr
                       g_R & = 1, \qquad \qquad v_R = 0.\cr }}
This is rather similar to \threetwist\ that gave us $N=5$ theory
except that there is no shift; as a result now there are additional
massless states in the twisted sector.
We get  $N=5$ supergravity from the untwisted
sector as in \threetwist.
To find the number of twisted sectors, note that $g_{L}$ leaves
$(0, p_R)$ invariant. So, the invariant lattice $I$ is the root lattice of
$SU(3)^3$, and $I^*$ is therefore the weight lattice of $SU(3)^3$.
The number of twisted sectors is $D\equiv \sqrt{{\rm det} (1-\theta_L)/
|I^*/I |}
=\sqrt{27/27} =1$. The single twisted sector contributes
an additional gravitino multiplet. Together,
we  obtain the gravity multiplet of
$N=6$ supersymmetry.

In this construction there is no obvious radial modulus to vary
so it is not clear  if this theory has a limit which gives the
$D=5, N=3$ theory and if so whether the limit is perturbative or
involves strong coupling.

The bosonic spectrum of $N=6$ supergravity
contains, in addition to  the graviton,
$32$ vector fields and $30$ scalars.
The symmetry group in this case is $SO^*(12)$. The vectors
transform in the $ 32$-dimensional spinor representation.
The scalars parameterize the coset $SO^*(12)/ U(6)$ \ADF.
Because the real rank of $SO^*(12)$ is three, there are
three non-compact moduli.

\subsec{$ D=5$}

\subsubsec{$N=2$}

We start with type II string theory at the point in Narain
moduli space defined by the lattice $\Gamma^{5,5}(D_5)$. The
Weyl group has a conjugacy class $D_5(a_1)$ of elements of
order $12$ with a pair of complex eigenvalues $\omega^2, \omega^3$
and a single real eigenvalue of $-1=\omega^6$. The shift vector
$v_R = (1,1,2,3,3)/12$ satisfies level matching and neither
$v_R$ nor $n v_R$, $n<12$ lie in $I^*$. This asymmetric orbifold thus
leads to pure $N=2$ supergravity in $D=5$.
The only scalar in the spectrum is the dilaton which parameterizes
the positive real line $\bR^+$.

\subsubsec{$N=3$}
This theory was discussed earlier as a perturbative large radius
limit of a $D=4,N=6$ theory.

\subsec{$ D=6$}

To obtain {$N= (1,1)$} pure supergravity,
we start with type II string theory at the point in Narain
moduli space defined by the lattice $\Gamma^{4,4}(A_4)$. The
Coxeter element of $A_4$ is order $5$ and has eigenvalues
$\omega, \omega^3$. This is in $SO(4)$ but not in $SU(2)$ so
twisting by this element on the left breaks all the left-moving
supersymmetries. The shift vector $v_R=(0,1,-1,2,-2)/5$ satisfies
level matching and is not in $I^*$ so there are no massless
states in the twisted sectors. The spectrum of this orbifold
is that of pure $N=(1,1)$ supergravity.
Again, the dilaton  parameterizes
the positive real line $\bR^+$.

\subsec{D=7 and 8}

These construction seems to fail when we get to seven or
eight dimensions because we cannot find appropriate shift
vectors. For $D=8$ we can give an exhaustive demonstration
that there are no asymmetric orbifold constructions of type
II string theory leading to the $(8,1,16)$ theory
in the low-energy limit.

We can
classify the possible asymmetric orbifolds
as follows. We start with a compactification on $T^2$ with moduli
space $ O(2,2,\bZ) \backslash O(2,2)/O(2)\times O(2)$,
or equivalently
\eqn\twomod{
[SL(2, \bZ)\backslash SL(2, \bR)/U(1) \times SL(2, \bZ)
\backslash SL(2,\bR)/U(1) ]/\bZ_2.}
We can twist by elements of the T-duality group if we are at a point
in the moduli space where these elements preserve the lattice and
therefore have fixed points acting on the above coset. Let $\sigma$ and
$\tau$ be the modular coordinates on each component of
\twomod. Then following \dvv\ we can write the $\Gamma^{2,2}$ lattice
using a complex basis as
\eqn\lbasi{\Gamma^{2,2} = {1 \over \sqrt{2 \Im \sigma \Im \tau}}
\bZ \pmatrix{1 \cr 1 \cr} \oplus \bZ \pmatrix{\bar \sigma \cr \sigma \cr}
\oplus \bZ \pmatrix{ \tau \cr \tau \cr} \oplus \bZ
\pmatrix{ \bar \sigma \tau \cr \sigma \tau \cr}}
Fixed points occur at
the orbifold point $\sigma=i, \tau=i$ which has an
enhanced $(Z_4 \times Z_4) \sdtimes Z_2$ symmetry,
at $\sigma=\rho$, $\tau=\rho$ with
$\rho=e^{2 \pi i/3}$ which has an enhanced $Z_3$ symmetry and at $\sigma=i$,
$\tau=\rho$ (or vice versa) which has a $Z_{12}$ symmetry which acts
quasicrystallographically \HMV\ as
\eqn\qcoact{\pmatrix{z_1 \cr z_2 \cr} \rightarrow \pmatrix{ -i \rho & 0 \cr
0 & i \rho \cr} \pmatrix{z_1 \cr z_2 \cr } }

The $Z_{12}$ twist breaks all the supersymmetry and so does not
yield the theory in question. Similarly, twists by a subgroup of
$Z_{12}$ at the point $\sigma=i$, $\tau=\rho$ also break all
the supersymmetry since they act both on the left and right.
At the point $\tau=\sigma=i$ the $Z_4$ or $Z_2$ abelian subgroups
of $(Z_4 \times Z_4)\sdtimes Z_2$ do not satisfy the mod two condition
\modtcon. At the
$Z_3$ symmetric point
one
can easily classify the possible shift vectors of order
three and find that shift vectors compatible with modular
invariance always lie in $I^*$ and thus lead to additional
massless fields in the twisted sector. We have not tried to
perform an exhaustive search of orientifold constructions or
heterotic constructions of this theory, but they seem unlikely
to exist. Note that no such constructions were found in the
detailed search of heterotic free fermion constructions carried
out in \CHL.

We have not tried to carry out a similar classification of possible
$D=7$ constructions, but the obvious possibilities all fail to
construct the $(7,1,16)$ theory.

It might be possible to obtain these theories in the context of
F-theory \VafaII. In \BKMT\ an extension of
F-theory is considered
where the monodromy of the coupling
constant field $\tau$ is in a subgroup of $SL(2, \bZ)$ because
of a nontrivial background of the 2-form $B_{\mu\nu}$ field.
Such compactifications also lead to vacua where some of the moduli
are frozen. More generally, it would be interesting to know whether
some or all of the string islands considered in this paper
can also be obtained as F-theory compactifications.

\newsec{General Comments and Speculations}

We have constructed all but two of the consistent pure supergravity
theories in $D \ge 4$ as the low-energy limit of asymmetric
orbifold string constructions.  Several of these theories have no
moduli other than the dilaton and we expect that they are self-dual
so that the strong coupling limit simply leads again to the same
theory. For the theories with $N_s=16$ in $D=4,5,6$ there is no
consistent higher dimensional theory with the right spectrum
that one could obtain by taking the strong coupling limit. The strong coupling
limit must therefore have the same low-energy spectrum, although
possibly with a different construction leading to a different massive
spectrum. For other cases such as the $(4,6,24)$ theory constructed
in 3.1.3.b the strong coupling limit is not clear. In all
these cases it would be interesting to study the non-perturbative
D-brane spectrum in order to determine the full U-duality group
and hence the strong coupling limit.

Finally, part of the motivation for this work was the possibility
of constructing lower-dimensional versions of M theory, that is
supersymmetric theories with no moduli at all which at low-energies
are described by one of the moduli-free pure supergravities $(5,1,8)$,
$(4,3,12)$, $(4,2,8)$ or $(4,1,4)$.  The $(5,1,8)$ theory in particular
has many similarities to M theory \refs{\cremmer, \chamnic, \mizohta}.
It is tempting to speculate that such theories might be obtained
as the strong coupling limit of theories in $D=4$ or $D=3$ which
have only the dilaton as a modulus. In particular, the $(5,1,8)$
theory could arise as the strong coupling limit of a $D=4$, $N=2$
theory with a single vector-multiplet. In fact, given such a theory,
this would seem to be the simplest candidate for the strong coupling limit.
Needless to say, we have not yet been able to construct such an
asymmetric orbifold although there does not seem to be any fundamental
reason why such a construction should not exist \DaHa. Of course,
given such a construction one would still need to exhibit
evidence for a Kaluza-Klein spectrum of soliton bound states.
For the time being this theory remains a ``Fantasy Island.''

\bigskip

\centerline{\bf Acknowledgements}\nobreak

It is a pleasure to thank
L. Dixon for helpful discussions and
for sharing some of his constructions, and S. Ferrara, A. Klemm,
G. Moore, M. S. Raghunathan and E. Silverstein for valuable discussions.
We would like to acknowledge the hospitality of the
Institute for Theoretical Physics at Santa Barbara,
and thank the organizers of the workshop on
`Duality in String Theory'  for inviting us.
A.~D. would like to acknowledge the hospitality
of the Enrico Fermi Institute at  the  University
of Chicago where some of this work was completed.
This work was  supported in part by the National Science Foundation
under Grant No.~PHY96-00697 and  Grant No.~PHY94-07194.

\listrefs
\bye

\end